\documentclass[a4paper,fleqn]{cas-dc}
\usepackage[numbers]{natbib}

\usepackage{tabularx}
\usepackage{booktabs}

\usepackage{tikz}
\usepackage{adjustbox}
\usetikzlibrary{positioning}

\usepackage{tcolorbox}

\usepackage{comment}

\def\tsc#1{\csdef{#1}{\textsc{\lowercase{#1}}\xspace}}
\tsc{WGM}
\tsc{QE}
\tsc{EP}
\tsc{PMS}
\tsc{BEC}
\tsc{DE}

\begin{document}
\let\WriteBookmarks\relax
\def\floatpagepagefraction{1}
\def\textpagefraction{.001}

\shorttitle{SoK: Post-Quantum Cryptography Implementation in Software}

\shortauthors{R.D.N. Shakya et~al.}

\title [mode = title] {SoK: Post-Quantum Cryptography Implementation in Software: Approaches, Challenges and the PQC-HOT Framework}

%
\author[UoM]{R.D.N. Shakya}[orcid=0000-0003-3803-273X]
\cormark[1]
\ead{shakyardn.26@uom.lk}
\credit{Investigation, Analysis, Writing}
\affiliation[UoM]{organization={Faculty of Information Technology, University of Moratuwa},
    addressline={Moratuwa}, 
    country={Sri-Lanka}}

\author[UoM]{C.P. Wijesiriwardana} [orcid=0000-0002-1124-425X]
\credit{Analysis, Supervision}
\ead{chaman@uom.lk}
\author[UoR]{S.M. Vidanagamachchi} [orcid=0000-0002-2245-4527]
\ead{smv@dcs.ruh.ac.lk}
\credit{Analysis, Supervision}
\affiliation[UoR]{organization={Faculty of Science, University of Ruhuna},
    addressline={Matara}, 
    country={Sri-Lanka}}
\author [RMIT] {Nalin A.G. Arachchilage} [orcid=0000-0002-0059-0376]
\ead{nalin.arachchilage@rmit.edu.au} 
\credit{Analysis, Supervision}
\affiliation[RMIT]{organization={School of Computing Technologies, RMIT University},
    addressline={Melbourne},  
    country={Australia}}

\cortext[cor1]{Corresponding author}

\nonumnote{This paper only used AI-based language editing to improve grammar and clarity. All intellectual, analytical, and interpretive work belongs to the authors.}


\begin{abstract}
Secure implementation of post-quantum cryptography (PQC) requires attention to cryptographic mechanisms, software integration, developer capability, and organisational support. Understanding how available approaches address these requirements is important for preparing software systems for quantum threats. This Systematisation of Knowledge (SoK) synthesises 30 publications and analyses PQC implementation approaches and challenges using a Human, Organisational, and Technological (HOT) perspective. We identify four approach categories: guidelines, frameworks, tools and libraries, and educational interventions. Technological support receives greater representation in the extracted mapping, while no approach is classified primarily as organisational, despite secondary organisational contributions in some approaches. The challenge synthesis identifies five layers covering implementation security, system integration and lifecycle, tooling, organisational governance, and human factors. Together, these findings highlight a difference between the primary support functions of the mapped approaches and the breadth of the reported implementation challenges. We propose PQC-HOT, an evidence-informed analytical framework connecting implementation tasks with technical resources, practitioner capabilities, and organisational arrangements. We derive research priorities and engineering implications to guide framework evaluation and support secure, maintainable PQC-enabled software.

\end{abstract}

\begin{keywords}
      Quantum Resilience \sep Post-Quantum Cryptography \sep PQC implementation \sep Software Engineering \sep HOT framework
\end{keywords}

\maketitle

\section{Introduction}
\label{intro}

Quantum computing (QC) represents a transformative advance in computational paradigms, utilising principles of quantum mechanics to surpass the limitations of classical computing \cite{bernstein_post-quantum_2017}. Its ability to solve computationally intractable problems at exceptional speeds has significant implications in computer science \cite{bernstein_post-quantum_2017, GILL20261}. In particular, QC introduces a fundamental disruption to existing cryptographic systems, as quantum algorithms can efficiently solve mathematical problems that underpin widely deployed encryption schemes \cite{dam_survey_2023}. For example, when a user logs into an online banking system, the connection is commonly protected using cryptographic schemes such as RSA (Rivest-Shamir-Adleman). These schemes rely on mathematical problems that are extremely difficult for today’s classical computers to solve within a practical time-frame. A classical computer might take several thousand years to break such encryption. However, quantum algorithms such as Shor’s algorithm could dramatically reduce this time, allowing a sufficiently powerful quantum computer to recover the secret encryption key and potentially expose sensitive information such as passwords or financial transactions \cite{bernstein_post-quantum_2017}. Similarly, Grover’s algorithm can significantly accelerate attacks against symmetric encryption by reducing the effort required to search for encryption keys \cite{bernstein_post-quantum_2017, dam_survey_2023}. Collectively, these developments indicate that emerging quantum capabilities pose a credible threat to widely adopted cryptographic standards, including RSA-2048, highlighting the urgent need for quantum resilience in software systems, the ability of systems to maintain secure and trustworthy operation despite quantum-enabled adversarial threats. In response, Post-Quantum Cryptography (PQC) has emerged as a key solution aimed at developing cryptographic schemes that remain secure against both classical and quantum adversaries \cite{bernstein_post-quantum_2017, dam_survey_2023}.

Although PQC algorithms provide strong theoretical security guarantees, their secure implementation in real-world software systems remains a significant challenge \cite{bernstein_post-quantum_2017}. In practice, cryptographic security depends not only on the strength of the underlying algorithm, but also on how well it is implemented, configured, and integrated into software \cite{crptoImmerr}. Prior studies have consistently demonstrated that secure cryptographic primitives can become vulnerable due to implementation flaws, insecure parametrisation, side-channel leakage, or incorrect usage by developers \cite{crptoImmerr, bagirovs_applications_2024, toruan2026security}. For example, weaknesses in key management or incorrect integration of cryptographic libraries can expose sensitive systems despite the use of theoretically secure algorithms \cite{crptoImmerr}. Therefore, achieving robust post-quantum security and long-term quantum resilience requires both sound cryptographic design and rigorously engineered implementation practices within software systems. 

These challenges are further intensified in the PQC context, where achieving quantum resilience requires balancing security, performance, compatibility, and implementation correctness. Compared to classical cryptographic primitives such as RSA, PQC schemes often introduce significantly larger key sizes, increased computational overhead, and higher memory requirements, which can stress existing software architectures and system constraints \cite{hekkala_implementing_2023, sodiya_quantum_2024}. Furthermore, PQC integration must occur within legacy systems that were not designed to accommodate such computational characteristics, creating compatibility and performance trade-offs that developers must carefully manage \cite{aydeger_towards_2024, giron_migrating_2023}. At the same time, tooling ecosystems, implementation standards, and developer expertise remain immature, limiting the availability of reliable guidance for secure implementation \cite{aydeger_towards_2024, kannwischer_improving_2022}. Consequently, software developers play a critical role in ensuring the security of PQC deployments, as incorrect configurations, insecure coding practices, or misunderstandings of PQC mechanisms can introduce vulnerabilities into otherwise quantum-resistant systems \cite{bagirovs_applications_2024, aydeger_towards_2024, giron_migrating_2023}.

Despite these practical concerns, existing research on PQC has focused primarily on algorithm design, security proofs, and performance benchmarking \cite{hekkala_implementing_2023}. While these contributions are essential, they provide limited insight for software developers into how PQC is implemented, deployed, and maintained within real-world systems. Consequently, current practice is often characterised by fragmented implementation, experimental integrations, and ad-hoc usage of cryptographic libraries, which increase the risk of misconfiguration and insecure deployments \cite{giron_migrating_2023}. In response, prior studies have proposed frameworks, guidelines, tooling support, and educational interventions to facilitate implementation \cite{hekkala_implementing_2023, kannwischer_improving_2022}. However, these efforts remain isolated, addressing narrow aspects of the problem without establishing a unified perspective on how technological, human, and organisational dimensions collectively shape PQC implementation outcomes.

This fragmentation suggests that achieving quantum resilience is not solely a cryptographic upgrade problem, but a broader challenge.  The successful implementation of PQC depends on the interaction between software architectures, developer practices, organisational governance, and operational requirements. However, existing studies typically examine these dimensions in isolation, resulting in a limited understanding of their interdependencies and combined impact on secure implementation. Consequently, the current body of knowledge lacks an integrated synthesis of PQC implementation approaches, their relationships, limitations, and the barriers that hinder secure and scalable deployment. This gap restricts both researchers and practitioners in developing a comprehensive understanding of the current implementation landscape and in identifying effective strategies for robust PQC implementation into real-world systems.

Therefore, a Systematisation of Knowledge (SoK) study is necessary to synthesise existing approaches, critically examine their limitations, and provide a structured foundation for advancing secure and practical PQC implementation toward quantum-resilient software systems. To systematically examine the current state of PQC implementation research, this study addresses the following research questions (RQs):

\begin{itemize} 
    \item RQ1: What methodologies, frameworks, guidelines, tools, and educational interventions have been proposed or developed to facilitate the implementation of Post-Quantum Cryptography (PQC) in software? 
    \item RQ2: What challenges and limitations arise when applying these methodologies, frameworks, guidelines, tools, and educational interventions to implement PQC in software?\end{itemize}

To address these RQs, we synthesise the reviewed literature using the Human, Organisational, and Technological (HOT) perspective~\cite{Berglund2020HTOA}. We connect the support functions of the approaches identified in RQ1 with the implementation challenges reported in RQ2. This synthesis informs PQC-HOT, an analytical framework for examining whether implementation tasks have corresponding technical resources, practitioner capabilities, and organisational arrangements. We illustrate its application to library integration and derive priorities for subsequent research and evaluation.

The study makes four contributions:

\begin{itemize}
    \item \textbf{A classification and HOT mapping of implementation approaches:} We organise the reviewed approaches into guidelines, frameworks, tools and libraries, and educational interventions, and map their primary HOT orientations. This mapping characterises the support represented in the extracted evidence set and identifies the absence of a primarily organisational intervention category.

    \item \textbf{A layered synthesis of implementation challenges:} We organise reported challenges into five layers: implementation security, system integration and lifecycle management, tooling and ecosystem support, organisational governance, and human factors. Relating these challenges to the approach mapping identifies questions about how available support addresses implementation requirements across the HOT dimensions.

    \item \textbf{The PQC-HOT analytical framework and an illustrative application:} We develop an evidence-informed framework connecting implementation tasks with technical resources, practitioner capabilities, and organisational arrangements. A hypothetical library-integration scenario illustrates how the framework can structure analysis across selection, integration, validation, and maintenance. Its proposed relationships and practical effectiveness remain subject to empirical evaluation.

    \item \textbf{A research agenda and engineering implications grounded in the synthesis:} We derive priorities for studying continuity across implementation activities, assessing developer support and capability, investigating organisational ownership of implementation assurance, and evaluating PQC-HOT. These priorities guide further investigation of the support required for secure and maintainable PQC implementation.
\end{itemize}

The remainder of this paper is organised as follows. Section \ref{relatedWork} reviews the related work. Section \ref{methodology} describes the research methodology. Section \ref{results} presents the results corresponding to the research questions. Section \ref{Discussion} provides the overall discussion, introduces the PQC-HOT framework, and outlines the implications of the research. Finally, Section \ref{conclu} presents the concluding remarks.

\section{Related Work}
\label{relatedWork}

The emergence of PQC has shifted attention from purely designing quantum-resistant algorithms toward understanding how these algorithms can be securely integrated into real-world software systems \cite{aydeger_towards_2024, BASERI2026104917}. Unlike classical cryptographic upgrades, PQC implementation introduces significant implementation challenges because many PQC schemes require larger key sizes, increased computational resources, and modifications to existing software architectures \cite{hekkala_implementing_2023, sodiya_quantum_2024}. Consequently, developers must balance security, performance, interoperability, and maintainability while integrating PQC into operational systems \cite{aydeger_towards_2024, giron_migrating_2023}. In response, researchers, standards bodies, and industry practitioners have proposed various forms of support, including frameworks, implementation guidelines, tooling ecosystems, and educational interventions intended to facilitate secure PQC implementation \cite{naether_migrating_2024, ahmed_survey_2025}.

Despite these efforts, current knowledge on PQC implementation remains fragmented across multiple research domains. Existing review studies often focus on isolated technological concerns such as algorithm benchmarking, protocol optimisation, or cryptographic library integration, while offering limited insight into how these components interact in practical software engineering environments \cite{naether_migrating_2024, ahmed_survey_2025}. For example, Ahmed et al. \cite{ahmed_survey_2025} examine the maturity and availability of PQC implementations in existing cryptographic libraries, yet provide limited discussion of the developer practices and organisational conditions required for secure deployment. Similarly, Näther et al. \cite{naether_migrating_2024} investigate migration processes and transition challenges, but primarily from an architectural and procedural perspective, with comparatively limited attention to developer usability, operational workflows, and governance considerations. Consequently, existing approaches frequently treat PQC implementation as a mainly technological upgrading problem, assuming that the availability of secure algorithms and implementation tools will naturally translate into secure and dependable software systems. 

However, cryptographic security in practice is strongly shaped by human, organisational and technological factors, including developer expertise, organisational processes, tooling maturity, and system integration constraints \cite{bagirovs_applications_2024, crptoImmerr, AdyaMishra_2022}. Cryptographic research demonstrates that vulnerabilities often emerge not from weaknesses in algorithms themselves, but from implementation errors, insecure configurations, and operational misuse \cite{crptoImmerr, toruan2026security, AdyaMishra_2022}. These risks become particularly significant in the context of PQC because many implementations remain immature, development ecosystems are still evolving, and practical deployment guidance is limited \cite{aydeger_towards_2024}. Consequently, understanding PQC implementation requires a broader perspective that considers how technological mechanisms interact with human and organisational factors within software development and deployment processes.

To address this gap, this paper presents a SoK on PQC implementation in software systems. Rather than examining algorithms or implementation tools in isolation, this study adopts a broader perspective to analyse how existing approaches collectively support or constrain secure PQC implementation in practice. The study synthesises research on existing approaches to PQC implementation in software, while critically examining their limitations and interdependencies. 

As shown in Table \ref{sok_Comparison}, previous review studies primarily emphasise technological or migration-centric perspectives. In contrast, this work provides a synthesis that integrates human, organisational, and technological dimensions. This perspective enables a deeper understanding of the structural challenges that hinder effective PQC implementation in real-world systems.

\begin{table*}[t]
    \centering
    \caption{Comparison of existing surveys and positioning of this SoK within the literature}
    \label{sok_Comparison}
    
    \resizebox{\textwidth}{!}{%
    \begin{tabularx}{\textwidth}{p{1.5cm} p{3cm} X}
        \toprule
        
        \textbf{Study} & \textbf{Scope and Focus} & \textbf{Remarks} \\
        \midrule
        
        Nadeem et al. \cite{ahmed_survey_2025} 
        & PQC cryptographic libraries and implementations 
        & \begin{itemize}
            \item Evaluates support, maturity, and availability of PQC algorithms in existing libraries
            \item Primarily focuses on technological implementation support and library availability, with limited discussion of developer practices, usability concerns, organisational readiness, or secure deployment processes
        \end{itemize}
        \\
        
        \midrule
        
        Näther et al. \cite{naether_migrating_2024} 
        & PQC migration processes 
        & \begin{itemize}
            \item Identifies migration phases, challenges, and transition strategies
            \item Emphasises migration workflows and process-level transition strategies, with comparatively limited attention to tooling ecosystems, governance mechanisms, and other non-technical factors
        \end{itemize}
        \\
        
        \midrule
        
        \textbf{This SoK} 
        & PQC implementation in software from a HOT perspective 
        & \begin{itemize}
            \item Systematises implementation approaches and analyses challenges across human–organisation–technology (HOT) dimensions
            \item Reveals cross-cutting challenges in PQC implementation
            \item Proposes the PQC-HOT framework
        \end{itemize}
        \\
        
        \bottomrule
    \end{tabularx}%
    }
\end{table*}

Furthermore, this SoK advances a multi-layered perspective on PQC implementation challenges by examining how HOT factors interact to influence security outcomes. By systematically organising and critically analysing the literature, this study establishes a structured foundation to guide future research and support the practical implementation of quantum-resilient software systems.

\section{Methodology}
\label{methodology}

A systematic literature review is conducted to identify PQC implementation approaches in software and their associated challenges. The study follows the two-stage protocol of Kitchenham and Charters, comprising planning and execution phases, to ensure a structured and reproducible review process \cite{kitchenham_procedures_2004}.

\subsection {Planning Stage}
\label{Planning_Stage}

In the planning stage, we define the review requirements and develop a structured protocol to minimise bias and ensure replicability \cite{kitchenham_procedures_2004}. We perform the following activities:

\begin{enumerate}
    \item Define the research scope
    \item Formulate the research questions
    \item Develop search strings
    \item Select data sources
    \item Define study selection criteria
\end{enumerate}

\subsubsection{Defining the Research Scope}
\label{Def_ResearchScope}

We define the research scope to guide the formulation of RQs, search strategies, and study boundaries \cite{booth_systematic_2016}. We apply the PICOC framework (Population, Intervention, Comparison, Outcome, Context) to structure the scope, as shown in Table \ref{picoc}. We then use this scope to guide the subsequent stages of the review.

\begin{table*}[ht]
    \centering
    \caption{Application of the PICOC Framework in this study}
    \label{picoc}
    \resizebox{\textwidth}{!}{%
    \begin{tabularx}{\textwidth}{p{2cm} X p{5cm}}
        \toprule
        
        Concept 
        & SLR Application 
        & Example Keywords \\
        \midrule
        
        Population 
        & Entities involved in implementing post-quantum cryptographic (PQC) mechanisms in software development 
        & developer*, secure software engineer* \\
        
        Intervention 
        & Methodologies, frameworks, guidelines, tools, or educational interventions used to implement PQC into software 
        & tool*, framework* \\
        
        Comparison 
        & \multicolumn{2}{l}{Not applicable as the SLR does not perform a comparison} \\
        
        Outcome 
        & Identification of available solutions (RQ1), understanding of challenges (RQ2) 
        & challenge*, limitation* \\
        
        Context 
        & Post-Quantum Cryptography implementation in software 
        & post-quantum cryptograph*, PQC \\
        
        \bottomrule
    \end{tabularx}}
\end{table*}

\subsubsection {Formulating the Research Questions}
\label{RQ_formulate}

Based on the defined scope, we examine existing approaches to implementing PQC \footnote{The scope of this study is limited to PQC implementation in software systems; approaches and challenges which are related to PQC migration are considered outside the scope.} in software and identify the associated challenges. Accordingly, we formulate the following Research Questions (RQs):

\begin{itemize}
    \item RQ1:  What methodologies, frameworks, guidelines, tools, and educational interventions have been proposed or developed to facilitate the implementation of Post-Quantum Cryptography (PQC) in software?
    
    \item RQ2: What challenges and limitations arise when applying these methodologies, frameworks, guidelines, tools, and educational interventions to implement PQC in software?
\end{itemize}

\noindent RQ1 investigates available interventions, while RQ2 examines their challenges and limitations. 

\subsubsection {Developing the Search Strings}
\label{Search_String}

\begin{figure*}[t] 
  \centering
  \includegraphics[width=\textwidth, height=6cm]{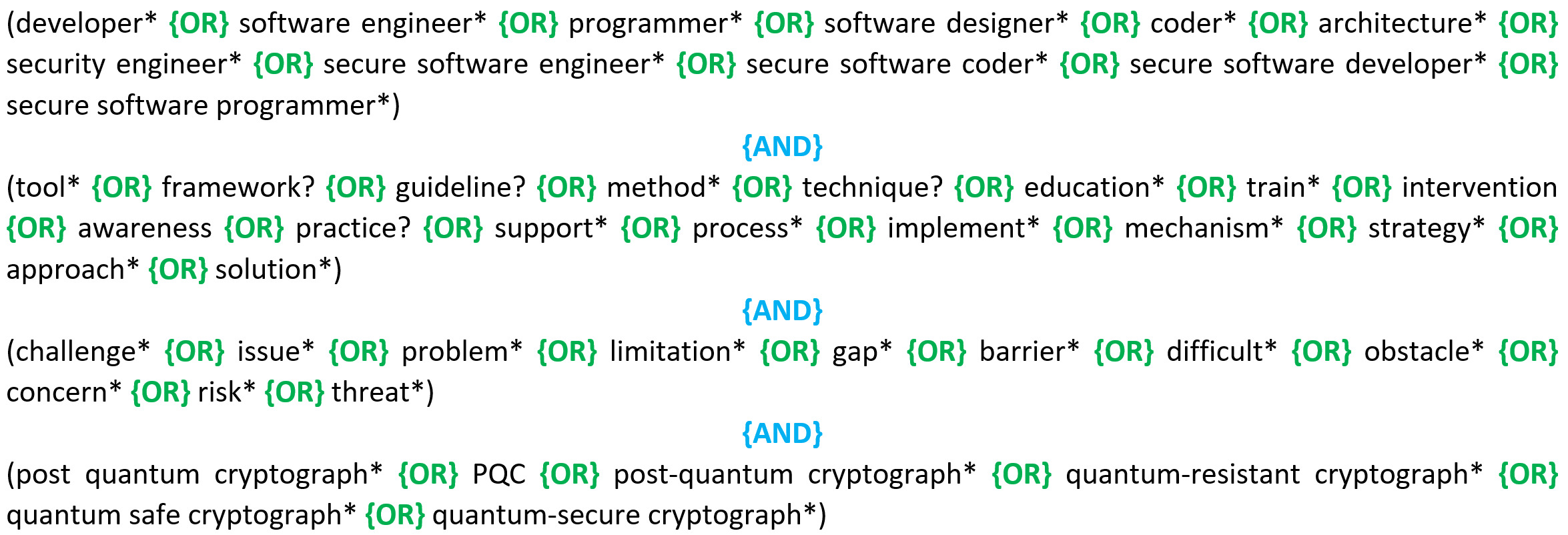}
  \caption{Constructed search string used for literature retrieval across selected digital libraries}
  \label{SearchString}
\end{figure*}

Then we developed a search string to retrieve literature relevant to this study's RQs. We derived keywords from the PICOC framework (Table \ref{picoc}) and constructed search strings using Boolean operators (AND, OR) to capture relevant variations and synonyms. Then, we applied the following search string across multiple digital libraries as illustrated in Figure \ref{SearchString}.

Due to database constraints, we divide the original search string into 17 sub-search strings to optimise retrieval.

\subsubsection {Selecting the Data Sources}
\label{data_Sources}

We retrieve literature from major scientific digital libraries, including ACM Digital Library, IEEE Xplore, Springer, Taylor \& Francis, and DBLP. We also conduct targeted searches in leading cybersecurity venues, including USENIX Security Symposium,  Symposium on Usable Security and Privacy (USEC), Journal of Computer Security, ACM Transactions on Computer-Human Interaction, Communications in Cryptology, ACM SIGSAC Conference on Computer and Communications Security (CCS), IEEE Transactions on Information Forensics and Security, and Transactions on Symmetric Cryptology, to ensure coverage of high-quality domain-specific research.

\subsubsection{Defining the Study Selection Criteria}
\label{studySelectCriteria}

Clear study selection criteria are essential to ensure the relevance and quality of an SLR \cite{booth_systematic_2016}. Accordingly, we formulated explicit inclusion (INC) and exclusion (EXC) criteria to guide the selection process, as outlined below:

\begin{itemize}
  \item \textbf{Inclusion Criteria}
      \begin{itemize}
       \item INC1: Research that explicitly addresses the implementation of PQC in software development.
        \item INC2: Studies that discuss methodologies, frameworks, guidelines, tools, or educational interventions related to PQC.
        \item INC3: Studies published in peer-reviewed journals or reputable conference proceedings.
        \item INC4: Literature published in English.
        \item INC5: Studies published between January 2020 and February 2026.\footnote{Two studies published in 2017 and 2019 were included as controlled exceptions through backward and forward snowballing due to their relevance. All other included studies strictly adhered to the defined inclusion window.}
      \end{itemize}

  \item \textbf{Exclusion Criteria}
      \begin{itemize}
        \item EXC1: Studies not focused on PQC implementation and lacking relevance to software development.
        \item EXC2: Non-peer-reviewed literature, such as blog posts, opinion pieces, white papers, or technical reports.
        \item EXC3: Duplicates or redundant publications presenting the same findings.
        \item EXC4: Articles without full-text access.
        \item EXC5: Non-English publications.
      \end{itemize}
\end{itemize}

\subsection {Conducting the Review Stage}
\label{Condcut_review}

After completing the planning phase, we perform the review following Kitchenham et al. \cite{kitchenham_procedures_2004}, consisting of study selection, data extraction, and data analysis.

\subsubsection{Study Selection} 
\label{Study_Selection}
\begin{figure*}[t] 
  \centering
  \includegraphics[width=\textwidth]{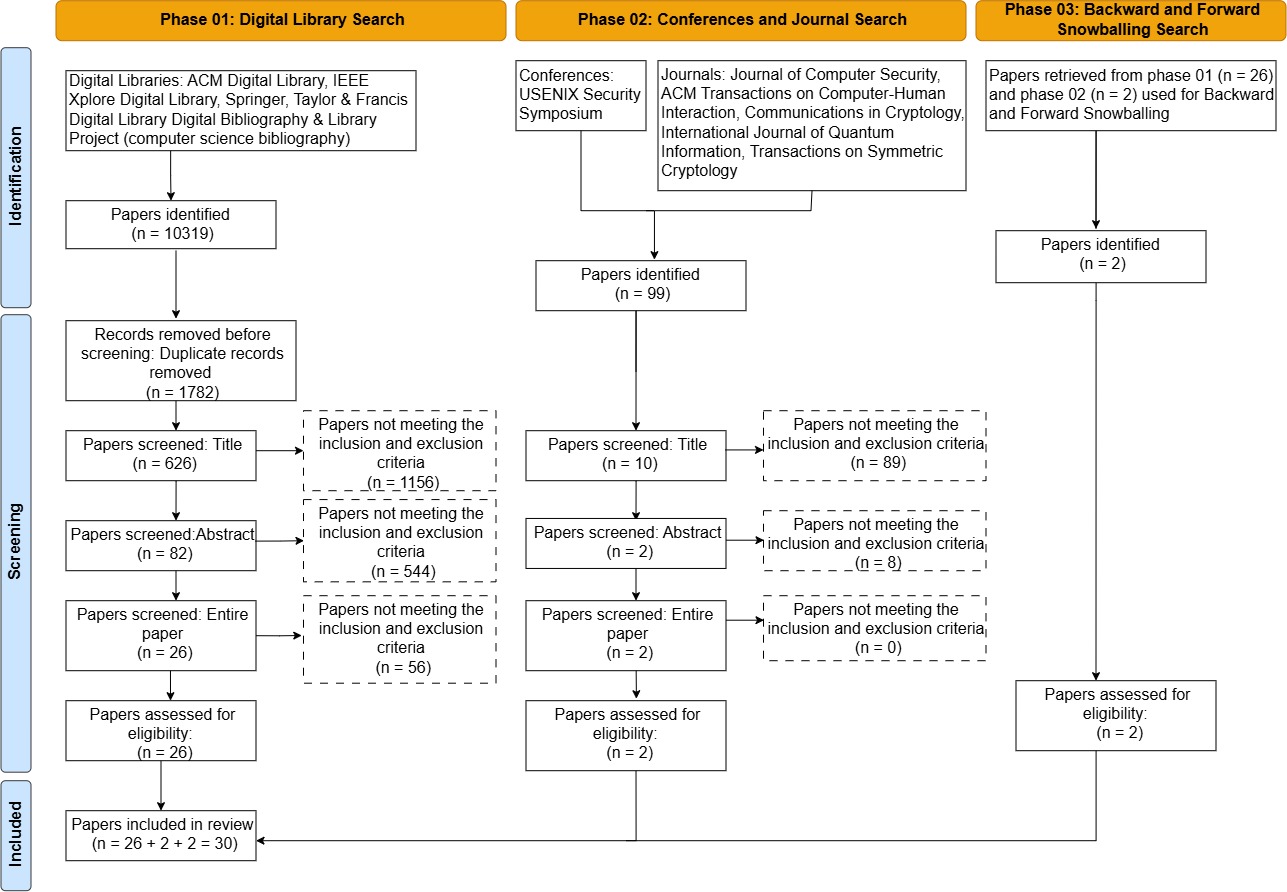}
  \caption{PRISMA flow diagram illustrating the study selection and screening process}
  \label{prisma}
\end{figure*}

We conducted the study selection process following the PRISMA 2020 framework \cite{page_prisma_2021}, as illustrated in Figure \ref{prisma}. 

We implemented this process in three structured phases. Each phase was carefully designed and conducted to ensure transparency and reproducibility in selecting the most relevant studies for inclusion in the SLR.

\begin{itemize}
    \item \textbf{Phase 1: Digital Library Search - }We systematically searched each digital library identified in Section \ref{data_Sources} using the search strings developed in Section \ref{Search_String}. 
    
    \item \textbf{Phase 2: Conference and Journal Search - }To mitigate possible publication bias \cite{kitchenham_procedures_2004}, we conducted a targeted search within the selected journals and conferences presented in Section \ref{data_Sources}, employing the search strings defined in Section \ref{Search_String}. 
    
    \item \textbf{Phase 3: Backward and Forward Snowballing Search - }To minimise the risk of omitting relevant studies, we performed both backward snowballing (reviewing the references cited in retrieved studies) and forward snowballing (identifying works that cite the retrieved studies) on the results obtained in Phases 1 and 2.
\end{itemize}

Phase 1 retrieved 10,319 publications. After removing duplicates and applying the inclusion and exclusion criteria through title, abstract, and full-text screening, 26 studies were selected. Phase 2 involved a targeted search of selected conferences and journals using the same 17 sub-search strings, yielding 99 publications. Screening identified 02 eligible studies. In Phase 3, we applied forward and backward snowballing to the studies selected in Phases 1 and 2. This resulted in the inclusion of 02 additional studies. Among these, two studies published in 2017 and 2019 were retained as exceptional cases due to their relevance. After completion of all three phases, we included a total of 30 articles in the systematic review.

\subsubsection{Data Extraction}
\label{Data_Extraction}

We extract both quantitative metadata (e.g., authors, year) and qualitative data (e.g., objectives, methods, findings). We use metadata for bibliographic organisation and qualitative data to address RQs. As the research focuses on software implementations of PQC, we exclude studies focused solely on hardware, IoT, or telecommunications domains.

\subsubsection {Data Analysis}
\label{Data_Analysis}

To address the RQs qualitatively, we applied the six-phase thematic analysis framework of Braun and Clarke \cite{braun_using_2006} to identify recurring themes in PQC implementation approaches and challenges. The analysis followed the six prescribed phases: 

\begin{enumerate}
    \item Data Familiarisation: We thoroughly reviewed the extracted text from the selected publications to gain an in-depth understanding of fundamental discussions on implementing PQC in software.
    
    \item Generating Initial Codes: We applied an integrated coding approach \cite{braun_using_2006}, combining deductive codes derived from the RQs and prior PQC knowledge with inductively generated codes emerging from the data. Codes were assigned to data segments relevant to the RQs. For example, the code “Testing and evaluation framework” was assigned to the statement: “we present PQClean, an extensive (continuous integration) testing framework for PQC software, ...” \cite{kannwischer_improving_2022}.
    
    \item Generating Themes: Codes were iteratively reviewed and grouped into broader themes representing recurring PQC implementation approaches and challenges, such as frameworks, tools, educational interventions, and limitations.
    
    \item Reviewing Themes: We refined these themes through iterative cross-checking and peer validation to ensure consistency, uniqueness, and alignment with the RQs, removing overlaps and ambiguities.
    
    \item Defining and Naming Themes: Each theme was clearly defined and assigned a concise descriptive label (e.g., “Frameworks” and “Tools and Libraries”).
    
    \item Producing the Report: Finally, we selected relevant extracts and combined the themes into a clear analytical narrative that directly addresses the RQs on examining approaches available for implementing PQC in software,  including the associated challenges.
\end{enumerate}

Inter-rater reliability was assessed using Cohen’s kappa \cite{irrCal} based on independent coding by co-authors. The results showed substantial agreement ($\kappa = 0.7945$). Differences were resolved through discussion and refinement of code definitions. 

\subsection{Human, Organisational, and Technological (HOT) Analysis}

To systematically synthesise the findings, we adopt the Human, Organisational, and Technological (HOT) framework \cite{Berglund2020HTOA}. The transition from classical cryptography to PC extends beyond the replacement of cryptographic algorithms. It represents a broader transformation that affects software engineering practices, developer competencies, organisational processes, and governance mechanisms \cite{crptoImmerr, bagirovs_applications_2024, AdyaMishra_2022}.

The HOT framework provides a suitable lens for analysing this multifaceted transition by categorising implementation factors into three interrelated dimensions. The \textit{Human (H)} dimension focuses on developer knowledge, skills, training, and support mechanisms that influence the secure implementation of PQC. The \textit{Organisational (O)} dimension includes governance, policies, planning, resource allocation, and organisational readiness that facilitate or constrain PQC implementation. The \textit{Technological (T)} dimension covers PQC algorithms, software libraries, implementation frameworks, engineering methodologies, testing approaches, and supporting tools required for integrating PQC into software.

Using this framework, the identified approaches and challenges reported in the literature were mapped to the corresponding HOT dimensions. This classification enables a structured synthesis of the current state of PQC implementation in software and provides a comprehensive perspective for addressing the RQs.

\subsection{Threats to Validity}

Following the guidelines of Kitchenham and Charters \cite{kitchenham_procedures_2004}, we acknowledge potential threats to validity across standard SLR dimensions.

\begin{itemize}
    \item \textbf{Search and Selection Validity:} We employ a systematic search strategy with explicit inclusion and exclusion criteria; however, we cannot eliminate the risk of omitting relevant studies. This limitation arises from database coverage constraints and indexing inconsistencies. Selection bias may also occur during screening. To mitigate these risks, we apply forward and backward snowballing, iteratively refine search queries, and document the selection process transparently.

    \item \textbf{Data Extraction and Researcher Bias:} Subjective interpretation may introduce bias during study selection, data extraction, and thematic synthesis. To reduce this risk, we follow a structured data extraction protocol, apply reflexive thematic analysis, and perform iterative cross-checking and peer validation to improve consistency and reliability. Inter-rater reliability was assessed using Cohen’s kappa \cite{irrCal}, which indicated substantial agreement ($\kappa = 0.7945$). All inconsistencies were resolved through discussion and refinement of code definitions.

    \item \textbf{Publication Bias:} By restricting the review to peer-reviewed, English-language publications, we introduce a bias toward well-established or positive results while excluding grey literature and non-English studies. This constraint may limit coverage and under-represent emerging or industry practices.

    \item \textbf{External Validity and Practical Applicability:} We synthesise findings from existing literature without independently validating them through expert or practitioner engagement. Although such validation could strengthen practical relevance, it falls outside the scope defined (refer to Table \ref{picoc}). Therefore, we present the findings as a structured synthesis of current knowledge rather than an empirical validation of practice.
\end{itemize}

\section{Results}
\label{results}

This section reports the synthesis of approaches supporting PQC implementation in software systems (RQ1) and the challenges described in the reviewed literature (RQ2). The approaches are organised by their implementation function and mapped to their primary Human, Organisational, or Technological (HOT) orientation. The challenges are grouped into five thematic layers, followed by cross-dimensional relationships identified through the synthesis. The findings characterise the reviewed evidence; they do not establish the prevalence or effectiveness of approaches across all PQC research and practice.

\subsection{RQ1: PQC Implementation Approaches in Software Systems}
\label{RQ1_Intervention}

\begin{figure*}[t]
  \centering
  \includegraphics[width=\textwidth,height=0.6\textheight,keepaspectratio]{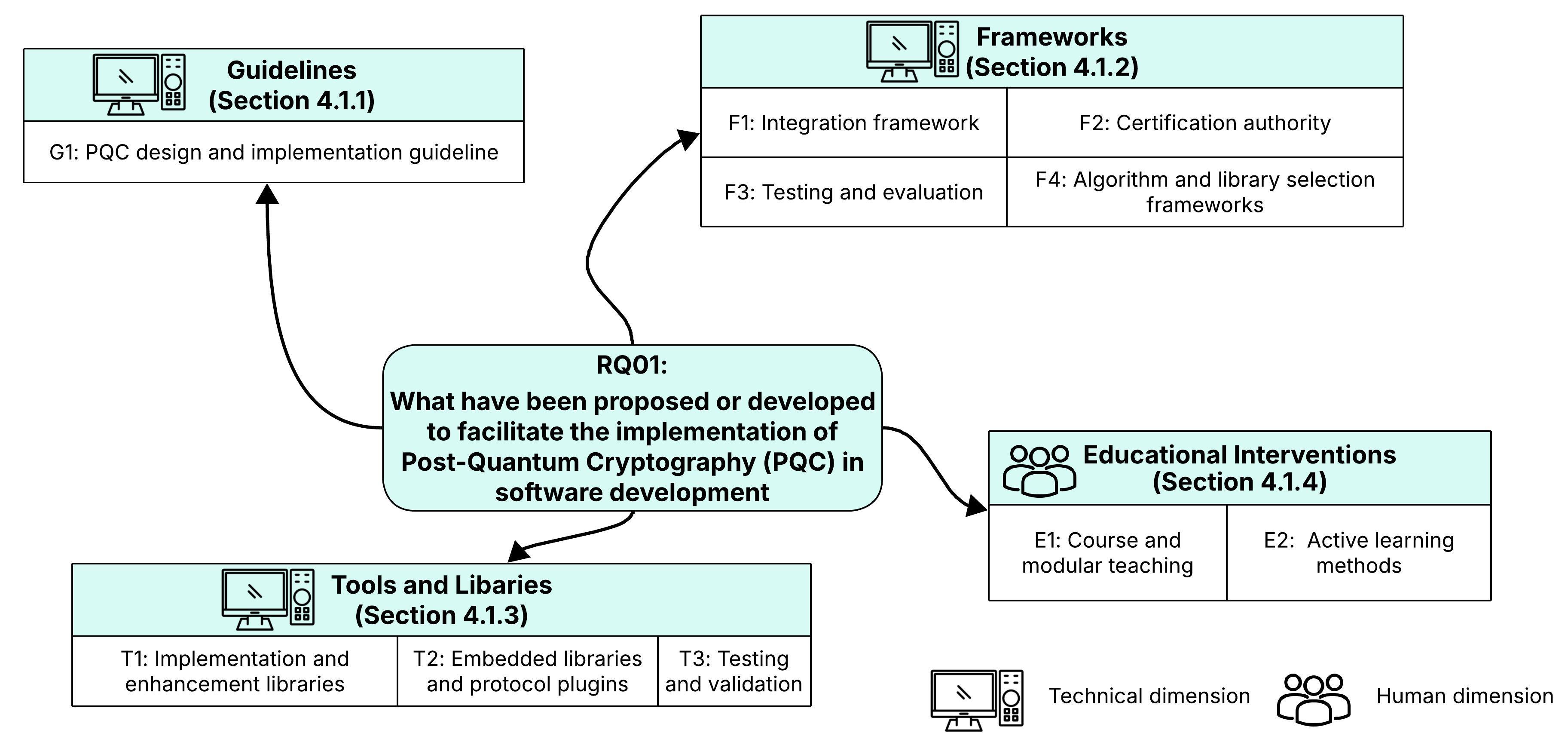}
  \caption{Overview of PQC implementation approaches identified in the literature (RQ1)}
  \label{RQ1Diagram}
\end{figure*}

The synthesis identified four approach categories: guidelines (G1), frameworks (F1-F4), tools and libraries (T1-T3), and educational interventions (E1-E2), as illustrated in Figure \ref{RQ1Diagram}. These categories describe different forms of support: design recommendations, structured integration and evaluation processes, executable implementation resources, and learning activities. 

\subsubsection{Guidelines (G1)}
\label{guide}

The PQC design and implementation guideline category (G1) primarily addresses the mathematical and cryptographic foundations of PQC algorithms, including formal security proofs, correctness guarantees, and hardness assumptions. Howe et al. \cite{howe_sok_2021} synthesise pitfalls in PQC constructions and implementations, highlighting the relationships among security assumptions, algorithm design, and implementation choices. For example, Learning with Errors (LWE) and Short Integer Solution (SIS) provide underlying hardness assumptions for lattice-based constructions. Such guidance supports practitioners in preserving the conditions required by a scheme’s security argument throughout design and implementation. Its primary orientation within the HOT framework is technological, as it addresses cryptographic mechanisms and their implementation; however, it does not constitute a complete software development or organisational adoption process.

\subsubsection{Frameworks (F1-F4)}
\label{frame}

Four framework subcategories organise support around integration, public key infrastructure, evaluation, and selection.

Integration frameworks (F1) describe architectures for incorporating PQC into software and communication systems. The reviewed frameworks either adopt hybrid cryptographic strategies that combine classical and PQC algorithms \cite{gulomov_exploring_2024, chandre_post-quantum_2024}, or propose novel architectures specifically designed for fully quantum-resilient systems \cite{ojha2025adoption}. 

Certification authority frameworks (F2) focus on extending existing public key infrastructure (PKI) models to support PQC algorithms in certificate generation, validation, and revocation.  These frameworks preserve trust relationships and authentication mechanisms in quantum-resilient environments while maintaining compatibility with evolving cryptographic standards \cite{tsili_scalable_2025}. 

Testing and evaluation frameworks (F3) emphasise the verification and assessment of PQC implementations. They provide mechanisms to evaluate implementation correctness, security, and performance, enabling developers and organisations to identify weaknesses and operational limitations before production integration \cite{kannwischer_improving_2022, e27020212, PQCLeo25}.

Algorithm and library selection frameworks (F4) assist developers in selecting the most appropriate PQC algorithms or software libraries based on implementation requirements. These frameworks support decision-making by evaluating multiple factors such as security level, computational performance, memory overhead, interoperability, and deployment constraints, thereby improving the suitability of PQC implementation \cite{decisionAlgo}.

These frameworks address different engineering decisions rather than a single common outcome. An architecture, a testing framework, and a selection method provide complementary support, but their classification does not establish that they are interoperable or collectively cover the implementation lifecycle.

\subsubsection{Tools and Libraries (T1–T3)}
\label{tool}

Tools and libraries for PQC implementation enable developers to implement, evaluate, test, and enhance PQC algorithms and their implementations within software systems. This category represents one of the most operationally significant groups of PQC approaches, as it directly supports implementation and deployment activities. The tools and libraries category comprises implementation and enhancement libraries (T1), embedded libraries and protocol integrations (T2), and testing and validation tools (T3). Across T1-T3, the primary HOT orientation is technological, although interfaces, documentation, and maintenance practices also have implications for developers and organisations.

T1 provides foundational support for developing and integrating PQC algorithms into software applications. These include hardness estimator libraries \cite{esser_sok_2024}, which assess the computational hardness and resilience of PQC schemes against both classical and quantum adversaries, as well as general-purpose PQC libraries that either provide standalone PQC implementations \cite{saucedo-estrada_post-quantum_2020} or extend existing cryptographic libraries with quantum-resistant capabilities \cite{hekkala_implementing_2023, ahmed_survey_2025}. Collectively, these libraries enable developers to experiment with, evaluate, and deploy PQC mechanisms in practical software environments.

T2 places PQC-related mechanisms within larger systems. Examples include the LaZer library for lattice-based zero-knowledge proofs \cite{lyubashevsky_lazer_2024}, PQConnect tunnels \cite{cryptoeprint_2024_2092}, post-quantum WireGuard \cite{huelsing2021wireguard}, and PQC integration with IPsec offloading \cite{aguilera_integrating_2024}. These examples demonstrate integration in specific settings; their inclusion does not establish suitability for every application or deployment environment.

T3 focuses on assessing the correctness, security, and performance of PQC implementations. These tools play a critical role in ensuring the reliability and trustworthiness of PQC-enabled systems by systematically evaluating quality and computational efficiency \cite{kannwischer_improving_2022}. For example, PQClean \cite{kannwischer_improving_2022} provides an example of software quality and continuous integration support for PQC implementations. Its inclusion under both F3 and T3 reflects its roles as a structured testing framework and a practical software resource. 

\subsubsection{Educational Interventions (E1–E2)}
\label{edu}

Educational initiatives play a critical role in enabling PQC implementation by addressing the human dimension. These interventions aim to transfer knowledge and develop the skills and capacities required to understand and implement PQC algorithms and principles. The educational category distinguishes course and modular teaching (E1) from active learning strategies (E2). 

Course and modular teaching approaches (E1) provide structured learning pathways to introduce PQC concepts in formal educational settings \cite{borrelli_towards_2025, borrelli_designing_2024}. These approaches primarily rely on course-based and modular instruction to establish the knowledge. They systematically introduce PQC principles, underlying cryptographic assumptions, and major algorithm families through structured teaching, enabling learners to build conceptual understanding and explore specific PQC techniques in depth \cite{borrelli_towards_2025, borrelli_designing_2024, edComScSociety, EducModule, PQCIP}.

In contrast, active learning strategies (E2) emphasise experiential, interactive, and learner-centred engagement. These approaches incorporate hands-on exercises, simulations, laboratory training, workshops, discussion forums, and feedback-driven activities to actively involve learners in the practical exploration of PQC concepts \cite{ abdelhamid_wip_2024}. Experiential learning methods provide direct interaction with PQC algorithms and implementations through practical exercises and simulated environments, strengthening conceptual understanding through application. Complementing these methods, interactive and collaborative approaches \cite{abdelhamid_wip_2024} promote peer learning and knowledge exchange through workshops, discussion-based activities, and collaborative problem-solving environments. Project-based learning further enables learners to engage with realistic PQC implementation scenarios and complex problem-solving tasks. Gamification techniques enhance learner motivation, accessibility, and sustained engagement by incorporating multimedia content, multiplayer environments, and extended reality frameworks into the learning process \cite{abdelhamid_wip_2024}. Additionally, AI-assisted adaptive learning content supports personalised learning by dynamically adapting educational content to learner needs \cite{PQCLabEdu}. For example, the Adaptive Lattice Parameter Tool (ALPT) applies reinforcement learning to provide an interactive PQC laboratory environment, enabling learners to adjust lattice parameters and observe real-time trade-offs between computational efficiency and security hardness. This approach enhances experiential learning by bridging theoretical PQC concepts with practical implementation \cite{PQCLabEdu}. Collectively, these educational interventions reflect a multidimensional effort to balance theoretical understanding, practical engagement, and learner motivation in advancing PQC awareness, knowledge, and implementation readiness.

The educational approaches primarily address the human dimension through knowledge development and learning activities. However, the presence of practical exercises, gamification, or adaptive features does not itself demonstrate secure implementation competence. The synthesis distinguishes the learning opportunities described from evidence that learners can integrate, test, and maintain PQC securely in software systems.

\subsubsection{Distribution Across the HOT Dimensions}
\label{RQ1_HOT}

Table \ref{sok_synthesis_hot} summarises the primary human, organisational, and technological (HOT) orientation of the extracted approach categories. 

\begin{table*}[t]
\centering
\caption{Primary HOT orientation of the approach categories in the extracted mapping}
\label{sok_synthesis_hot}

    \begin{tabularx}{\textwidth}{p{3cm} p{5cm} X c}
        \toprule
        \textbf{Dimension} & \textbf{Intervention Type} & \textbf{Mapped Publications} & \textbf{Count} \\
                
        \midrule
        
        \textbf{Technological} 
        & Frameworks (F1-F4) 
        & \cite{kannwischer_improving_2022, gulomov_exploring_2024, chandre_post-quantum_2024, ojha2025adoption, tsili_scalable_2025, e27020212, PQCLeo25, decisionAlgo}
        & 8 \\
        
        & Guidelines (G1) 
        & \cite{howe_sok_2021}
        & 1 \\
        
        & Tools/Libraries (T1-T3) 
        & \cite{hekkala_implementing_2023, kannwischer_improving_2022, ahmed_survey_2025, esser_sok_2024, saucedo-estrada_post-quantum_2020,lyubashevsky_lazer_2024, cryptoeprint_2024_2092, huelsing2021wireguard, aguilera_integrating_2024}
        & 9 \\
        
        \midrule
        
        \textbf{Human} 
        & Educational Interventions (E1-E2) 
        & \cite{borrelli_towards_2025, borrelli_designing_2024, edComScSociety, EducModule, PQCIP, abdelhamid_wip_2024, PQCLabEdu}
        & 7 \\

        \midrule
        
        \textbf{Organisational} 
        & - 
        & -
        & 0 \\
        
        \bottomrule
    \end{tabularx}
\par\smallskip
\begin{minipage}{\textwidth}
\footnotesize
\textit{Note:} Counts denote publications assigned to each approach category, not independent interventions or effectiveness estimates. PQClean \cite{kannwischer_improving_2022} appears under both frameworks and tools/libraries. The rows therefore contain 26 assignments representing 25 distinct publications. HOT orientation denotes the primary support function; it does not exclude secondary contributions to other dimensions. The educational count retains the extracted classification and requires confirmation of PQC-specific eligibility.
\end{minipage}
\end{table*}

Technologically oriented approaches accounted for the largest share of the mapping, comprising 17 distinct publications across frameworks (F1-F4; eight assignments), guidelines (G1; one assignment), and tools/libraries (T1-T3; nine assignments). The human dimension comprised ten publications classified as educational interventions (E1-E2), subject to confirmation of their PQC-specific eligibility. Thus, the
extracted mapping contained more publications primarily oriented towards technological support than towards human capability development.

No approach was classified primarily within the organisational dimension. This finding indicates that the mapping identified no dedicated organisational intervention, such as a governance process or organisational capability-development approach. It does not establish that organisational considerations were absent from the reviewed publications. Organisational barriers were identified in the RQ2 synthesis, indicating a contrast between the challenges reported and the primary support functions of the approaches mapped.

Overall, the HOT distribution shows greater representation of technological support, a smaller body of educational interventions, and no primarily organisational interventions within the extracted evidence set \footnote{These counts describe the distribution of mapped publications; they do not establish methodological quality, industrial adoption, comparative effectiveness, or the extent of integration between approaches.}. 

\subsection{RQ2: Challenges in PQC Implementation}
\label{RQ2_challenges}

\begin{figure*}[t]
  \centering
  \includegraphics[width=\textwidth,height=0.8\textheight,keepaspectratio]{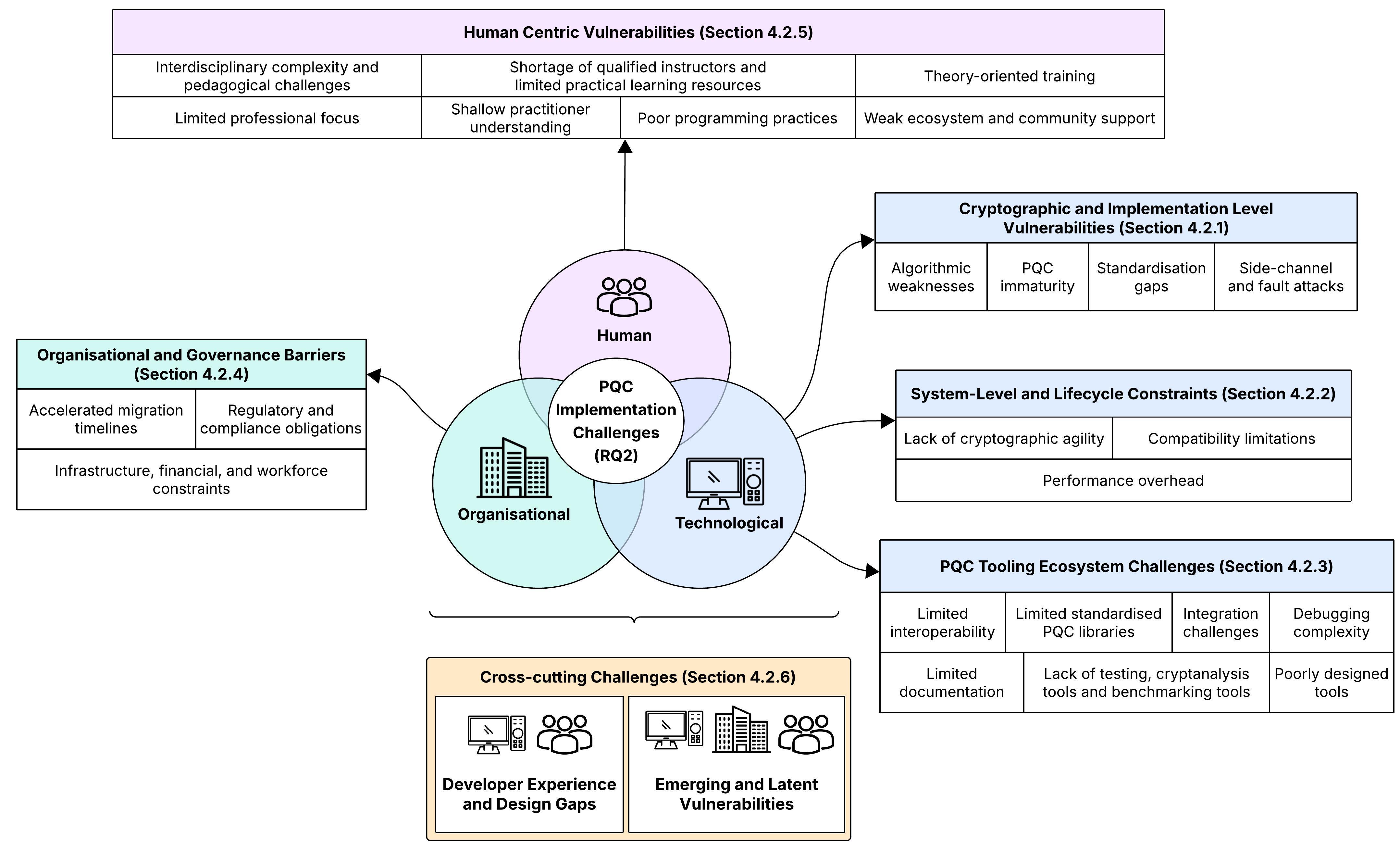}
  \caption{Challenges in PQC implementation in software systems (RQ2), structured across Human, Organisational, and Technological dimensions}
  \label{HOT_RQ2}
\end{figure*}

The reviewed literature describes challenges at five levels: cryptographic and implementation level; system-level and lifecycle; tooling and ecosystem; organisational and governance; and human-centred. Figure \ref{HOT_RQ2} organises these challenges across the HOT dimensions. The first three layers primarily concern technology, while the remaining layers foreground organisational and human factors. These groupings identify the focus of a challenge \footnote{The evidence does not support a quantitative ranking of their frequency, severity, or contribution to implementation failure.} rather than mutually exclusive causes. 

\subsubsection {Cryptographic and Implementation Level Vulnerabilities}
\label{F_L01}

PQC introduces significant security challenges at the intersection of cryptographic design and software implementation, because its theoretical guarantees depend heavily on correct and reliable implementation \cite{zhang_making_2023}. Even mathematically secure algorithms can become vulnerable when weaknesses emerge in their practical deployment \cite{crptoImmerr}. As a result, vulnerabilities often arise not only from flaws in cryptographic constructions themselves, but also from implementation decisions, optimisation techniques, parameter choices, and engineering practices \cite{howe_sok_2021}. 

One major concern is the presence of algorithmic and structural weaknesses that adversaries can exploit to recover sensitive information \cite{howe_sok_2021}. For example, re-creation attacks leverage publicly available parameters and structural patterns in cryptographic operations to reconstruct private keys, particularly in noisy constructions such as lattice-based and code-based schemes \cite{howe_sok_2021}. Security-critical operations, including key generation, encapsulation, and decryption, therefore represent high-risk attack surfaces. Any deviation from correct implementation, such as leakage through intermediate states or improper exposure of sensitive operations, can directly compromise system security \cite{howe_sok_2021}.

These risks are amplified by the relative immaturity of many PQC algorithms and implementations \cite{bagirovs_applications_2024, aydeger_towards_2024, ojha2025adoption}. Compared with well-established classical cryptographic primitives, many PQC schemes remain newly developed, insufficiently tested, or not yet fully standardised. This lack of maturity creates uncertainty regarding their robustness under diverse real-world deployment conditions and increases the possibility of unforeseen vulnerabilities emerging over time \cite{bagirovs_applications_2024}. 

In addition, limitations in current standardisation outputs further contribute to implementation-level risks \cite{aydeger_towards_2024, kannwischer_improving_2022, ojha2025adoption}. Although initiatives such as the NIST PQC standardisation process have identified mathematically secure algorithms, their reference implementations frequently lack robustness, portability, and adherence to secure software engineering practices \cite{aydeger_towards_2024, kannwischer_improving_2022, ojha2025adoption}. Many implementations remain architecture-specific, fragile, and susceptible to memory safety issues such as buffer overflows. Consequently, developers are often required to bridge the gap between theoretical specifications and deployment-ready secure implementations, increasing the probability of introducing implementation flaws. 

Despite their mathematical strength, PQC schemes therefore remain highly susceptible to exploitation at the implementation level for side-channel and fault attacks \cite{hekkala_implementing_2023, howe_sok_2021}. Developers may incorrectly apply transformations such as Fujisaki--Okamoto, mishandle memory operations, introduce unsafe optimisations, or fail to eliminate timing inconsistencies, all of which can undermine formal security guarantees \cite{hekkala_implementing_2023, howe_sok_2021}. For instance, attackers may exploit sparsity in code-based schemes or weaknesses in algorithms such as BIKE to recover cryptographic keys when implementations fail to enforce strict operational controls \cite{howe_sok_2021}. Furthermore, PQC implementations are particularly vulnerable to side-channel and fault-based attacks \cite{hekkala_implementing_2023, howe_sok_2021}. Unlike attacks targeting cryptographic design itself, these attacks exploit physical or execution-level leakage generated during cryptographic operations. Techniques such as differential power analysis, cold-boot attacks, and fault injection can reveal sensitive information through observable execution patterns \cite{hekkala_implementing_2023, howe_sok_2021}. Computationally intensive operations, including syndrome decoding, further increase this exposure, especially when implementations do not employ constant-time execution and leakage-resistant techniques \cite{hekkala_implementing_2023, howe_sok_2021}. 

Collectively, these challenges demonstrate that PQC security depends not only on the mathematical strength of cryptographic algorithms but also on the quality and security of their implementation. Vulnerabilities frequently emerge at the boundary between cryptographic theory and software engineering, where weaknesses in coding practices, optimisation strategies, side-channel resistance, and implementation standardisation can undermine formal security guarantees. Consequently, secure PQC deployment requires rigorous secure software engineering practices, comprehensive implementation validation, constant-time design principles, and continuous vulnerability assessment to ensure resilience in real-world environments.

\subsubsection{System-level and Lifecycle Constraints}
\label{F_L02}

The implementation of PQC introduces challenges that extend beyond individual cryptographic components and affect the entire software lifecycle. These challenges arise because the requirements of PQC frequently conflict with the architectural assumptions, operational constraints, and long-established design practices of existing software systems. As a result, organisations must address not only algorithmic integration but also broader issues related to system architecture, planning, interoperability, and long-term maintenance.

A major challenge is the lack of cryptographic agility within many existing systems \cite{tsili_scalable_2025, bavdekar_post_2023}. In numerous software architectures, cryptographic functions are tightly coupled with application logic, communication protocols, and data-processing workflows, making the replacement or introduction of new algorithms difficult \cite{crypotAgileCouppling}. Because PQC standards and recommendations continue to evolve, developers are often required to redesign substantial portions of a system rather than perform incremental updates. This significantly increases implementation effort, development cost, and the possibility of introducing new vulnerabilities or operational errors during system modification \cite{agilWork}.

These difficulties are further complicated by compatibility limitations across legacy systems and heterogeneous infrastructures. Many existing platforms were not designed to support the computational demands, interface requirements, or architectural flexibility associated with PQC algorithms \cite{aydeger_towards_2024, giron_migrating_2023, chandre_post-quantum_2024, ojha2025adoption, bavdekar_post_2023}. Consequently, organisations frequently rely on intermediary abstraction layers or hybrid cryptographic deployments that combine classical and post-quantum algorithms. This increases architectural complexity and expands the risk of misconfiguration, integration failures, and operational instability.

In addition, PQC introduces significant performance and resource overhead. Compared with many classical cryptographic schemes, PQC algorithms typically require larger keys, increased bandwidth, higher memory consumption, and greater computational resources \cite{bagirovs_applications_2024, aydeger_towards_2024, giron_migrating_2023, gulomov_exploring_2024, chandre_post-quantum_2024, ojha2025adoption}. These overheads can reduce system efficiency, increase latency, and complicate scalability. The impact is particularly severe in resource-constrained environments such as embedded systems and IoT platforms, where processing power, storage capacity, and energy availability remain limited \cite{giron_migrating_2023, chandre_post-quantum_2024, ojha2025adoption}. As a result, maintaining an effective balance between security, interoperability, and performance remains a major implementation challenge.

Overall, these limitations demonstrate that PQC implementation represents a broader system-level and lifecycle-oriented challenge rather than an isolated cryptographic concern. The associated risks emerge not only from individual technological components but also from the interaction between architectural rigidity, interoperability demands, performance limitations, and long-term operational requirements. Consequently, successful PQC implementation requires lifecycle-aware system design, continuous implementation, strategic planning, and sustained risk management rather than a one-time technological upgrade. 

\subsubsection{PQC Tooling and Ecosystem Challenges} 
\label{F_L03}

The PQC tooling ecosystem introduces risks that directly affect implementation security, reliability, and developer productivity. 

A fundamental challenge emerges from the fragmented nature of the PQC ecosystem, where tools, libraries, and frameworks often provide limited interoperability \cite{aydeger_towards_2024}. Developers frequently combine multiple independent components for implementation, testing, and validation, increasing the probability of inconsistent configurations, integration errors, and insecure deployment practices. This fragmentation reflects a broader misalignment between cryptographic innovation and software engineering practice.

These issues are further intensified by the scarcity of well-maintained standardised PQC libraries \cite{hekkala_implementing_2023, aydeger_towards_2024, ahmed_survey_2025}. Many available libraries remain experimental, insufficiently validated, or poorly maintained, creating uncertainty about correctness, portability, and long-term security guarantees. Consequently, developers often struggle to identify reliable implementation artefacts suitable for production environments.

A related challenge arises when developers try to integrate PQC mechanisms into existing development, security, and networking infrastructures, such as compilers, communication frameworks, and VPN systems \cite{hekkala_implementing_2023}. Translating mathematically intensive PQC algorithms into robust and performant software systems frequently requires considerable reworking across software stacks, development workflows, and deployment environments. As a result, integration complexity increases significantly, particularly in heterogeneous and legacy infrastructures.

Verification and validation processes also remain underdeveloped. The complexity and performance sensitivity of PQC algorithms make debugging inherently difficult \cite{hekkala_implementing_2023}, particularly when implementations require optimisation for performance and side-channel resistance. Limited documentation \cite{toruan2026security, hekkala_implementing_2023} and insufficient testing mechanisms and lack of advanced cryptographic analysis and benchmarking tools \cite{bagirovs_applications_2024, hekkala_implementing_2023, ojha2025adoption, PQCLeo25, aguilera_integrating_2024, rattanavipanon_toolchain_2025} further make it difficult for developers to establish implementation correctness and security assurance before deployment. Consequently, latent vulnerabilities may remain undetected until systems are operational.

Poorly designed tooling can introduce insecure defaults, expose unsafe low-level abstractions, omit comprehensive validation mechanisms, and unintentionally facilitate vulnerabilities within the development pipeline \cite{toruan2026security, hekkala_implementing_2023}. As a result, developers may incorrectly configure cryptographic parameters, misuse APIs, or adopt unsafe coding practices during implementation. These issues collectively increase the possibility of implementation errors, undermine correctness, and ultimately weaken the overall security posture of software systems \cite{toruan2026security, hekkala_implementing_2023}.

Critically, tools act as risk multipliers. Therefore, tooling does not merely support implementation; it fundamentally shapes the security properties of the resulting system. The PQC tooling landscape thus represents not simply a supporting infrastructure, but a critical challenge layer where ecosystem deficiencies can systematically undermine the security, reliability, and trustworthiness of post-quantum deployments.

\subsubsection{Organisational and Governance Barriers}
\label{F_L04}

Organisational factors represent a critical yet often overlooked dimension of PQC implementation. Although much of the current discourse focuses on cryptographic algorithms and technological integration, secure and sustainable implementation also depends heavily on organisational readiness, governance structures, strategic planning, and resource management \cite{orgReadiness}. Technological solutions alone are insufficient if organisations lack the institutional capacity required to manage, enforce security practices, and sustain long-term operational implementation.

The organisational governance challenge is heavily affected by uncertainty regarding the future progression of quantum computing development. Since the pace of advancement remains difficult to predict, organisations face challenges in determining appropriate implementation timelines, with the risk that sudden breakthroughs may compress available response windows and require rapid, potentially suboptimal decisions \cite{bavdekar_post_2023}. In parallel, the “record now, decrypt later” threat model significantly accelerates the urgency of implementation \cite{giron_migrating_2023, e27020212}. Adversaries may capture encrypted data today with the intention of decrypting it in the future once sufficiently powerful quantum systems become available \cite{giron_migrating_2023, e27020212}. This long-term confidentiality risk pressures organisations to implement PQC solutions in advance of full ecosystem maturity, including stable standards, robust implementations, and well-established operational practices. Consequently, implementation efforts are often undertaken under constrained timelines, increasing the possibility of premature or suboptimal implementation decisions.
 
In parallel, regulatory and compliance obligations introduce additional operational complexity into PQC implementation. Organisations implementing PQC must continue to satisfy legal frameworks, data protection regulations, and certification requirements. These challenges become particularly complex in hybrid environments where classical and PQC mechanisms coexist \cite{aydeger_towards_2024}. As a result, organisations must simultaneously manage interoperability, compliance assurance, and evolving security requirements across heterogeneous infrastructures. These obligations complicate deployment planning, increase operational overhead, and constrain implementation flexibility. Furthermore, different regulations across jurisdictions can lead to inconsistent deployment practices, additional compliance work, and delays in organisational adaptation \cite{ojha2025adoption}.

From an infrastructure perspective, PQC implementation frequently requires considerable upgrades to computational resources, storage systems, and communication infrastructures because PQC algorithms generally force greater computational and communication overhead than many classical cryptographic schemes \cite{aydeger_towards_2024}. These infrastructure demands create significant financial pressures and resource allocation challenges. Consequently, planning often extends beyond software modification to include broader infrastructure modernisation efforts. Human capital and expertise limitations represent another major organisational barrier \cite{aydeger_towards_2024, PQCIP}. Effective PQC implementation requires expertise across cryptography, secure software engineering, systems integration, and cybersecurity governance. However, many organisations currently face shortages of professionals with the needed interdisciplinary skills. As a result, organisations must invest in workforce training, capability development, and long-term knowledge retention, all of which increase operational costs and limit scalability.

Overall, organisational capability emerges as a critical factor in the successful implementation of PQC systems. The challenges associated with governance, compliance, infrastructure readiness, workforce capability, and strategic planning demonstrate that PQC implementation is as much an organisational transformation problem as it is a technological one. Without robust governance structures, standardised processes, sustained investment, and proper planning, even technologically secure PQC solutions may fail to achieve reliable, scalable, and sustainable deployment. Consequently, organisational readiness ultimately determines whether PQC systems remain theoretical security mechanisms or evolve into operationally resilient real-world deployments. 

\subsubsection{Human Centric Vulnerabilities}
\label{F_L05}

Human factors introduce a universal and often underestimated challenge in PQC implementation. Developers act as the final layer in the security pipeline, where gaps in knowledge, training, and ecosystem support translate directly into vulnerabilities \cite{devFault}. These challenges emerge from both educational deficiencies and limitations in practical implementation support, creating a disconnect between theoretical understanding and real-world deployment capability.

Current PQC education and awareness initiatives remain limited primarily due to the interdisciplinary nature of the field. Since PQC spans cryptography, software engineering, and quantum computing, institutions often struggle to provide sufficient interdisciplinary coverage and implementation-focused learning opportunities \cite{borrelli_designing_2024}. By this, educators also face pedagogical challenges related to determining appropriate technical depth \cite{borrelli_designing_2024}, managing interdisciplinary content \cite{borrelli_designing_2024}, and integrating active learning approaches such as hands-on exercises and gamified learning strategies \cite{abdelhamid_wip_2024}. Consequently, delivering balanced and practice-oriented PQC education remains difficult across both academic and professional contexts.

This challenge is further increased by the shortage of qualified instructors capable of teaching across these domains  \cite{borrelli_designing_2024}. As a result, the development of a skilled and implementation-ready workforce remains slow. The lack of accessible practical teaching resources further reduces the effectiveness and accessibility of PQC education \cite{borrelli_towards_2025, borrelli_designing_2024}. 

Most existing educational initiatives maintain a strong theoretical focus and primarily target university or high school students with technical backgrounds \cite{borrelli_towards_2025, borrelli_designing_2024, EducModule, abdelhamid_wip_2024}. These initiatives rarely address the practical needs of professional developers or real-world implementation environments \cite{borrelli_towards_2025, borrelli_designing_2024, abdelhamid_wip_2024}, although a small number of emerging approaches introduce limited laboratory-based exercises \cite{PQCLabEdu} or practitioner-oriented \cite{PQCIP} modules. However, such interventions remain fragmented, often scoped to isolated experimental settings, and are not yet integrated into comprehensive curricula or widely adopted training pathways. Practice-oriented professional training, particularly hands-on laboratories simulating integration, debugging, performance optimisation, and constant-time implementation challenges, therefore, remains underdeveloped at scale. In addition, targeted interventions for professional developers are scarce and typically lack longitudinal evaluation or industrial validation \cite{PQCIP}. Consequently, despite isolated progress, many learners continue to acquire primarily conceptual understanding of PQC without consistently developing the practical competencies required for secure deployment, and current educational efforts remain insufficient in producing implementation-ready practitioners.

Furthermore, developers, system administrators, and decision-makers often have only a shallow understanding of PQC principles, algorithms, and secure usage patterns \cite{giron_migrating_2023}. This limited expertise increases the possibility of misconfiguration, incorrect API usage, insecure integration practices, implementation errors, and poor programming practices. In many cases, practitioners also fail to apply secure software engineering principles specifically required for PQC implementation, such as constant-time execution, safe parameter selection, memory-safe programming, and leakage-resistant implementation techniques \cite{toruan2026security, hekkala_implementing_2023, howe_sok_2021}. Isochrony-related issues, including timing inconsistencies introduced through unsafe coding and optimisation practices, further increase the risk of side-channel leakage and exploitable vulnerabilities \cite{hekkala_implementing_2023, howe_sok_2021}. Consequently, even mathematically secure algorithms may become vulnerable when implemented incorrectly.

Weak ecosystem support further amplifies these risks. Existing libraries, documentation, and vendor or community support remain insufficiently mature and accessible, which complicates implementation, troubleshooting, and validation activities \cite{toruan2026security, giron_migrating_2023}.

Collectively, these factors establish human resources as a critical challenge in PQC implementation. Cognitive limitations, insufficient practical training, weak ecosystem support, and insufficient implementation guidance can transform the human layer into a critical point of failure, where the theoretical security guarantees of PQC are undermined during real-world implementation and deployment.

\subsubsection{Cross-Dimensional Challenges}
\label{RQ2Cross}

The analysis reveals that several PQC implementation challenges are not restricted to isolated dimensions but instead emerge through interactions among Human, Organisational, and Technological (HOT) dimensions. These challenges exhibit HOT coupling, where limitations in one aspect influence behaviours, decisions, and outcomes in others, producing compounded implementation risks. Accordingly, PQC implementation cannot be sufficiently understood through isolated technological, human, or organisational perspectives, as weaknesses propagate across these interacting domains and amplify system-wide vulnerability. This interdependence indicates that mitigation strategies must address the broader implementation ecosystem rather than individual components in isolation.

The analysis further identifies several cross-cutting challenges that arise through these interdependencies. These should be interpreted not as discrete or exhaustive categories, but as recurring interaction patterns that reflect structural coupling within PQC implementation environments.

\begin{itemize}
    \item \textbf{Developer Experience and Design Gaps:} \label{F_Cr_01}
\end{itemize}

The PQC ecosystem demonstrates persistent limitations in developer experience and in the design of implementation toolchains. While existing research prioritises cryptographic correctness and formal security guarantees, comparatively less attention is given to usability \footnote{In this context, usability refers to developer-facing characteristics, including API clarity, abstraction suitability, secure-by-default design, developer experience, and the ease with which cryptographic primitives can be correctly integrated and embedded into software systems \cite{toruan2026security}.}, integration pathways, and implementation ergonomics for software engineers \cite{toruan2026security}. Current implementations frequently expose low-level cryptographic interfaces, fragmented APIs, and domain-specific terminology \cite{toruan2026security}. These characteristics increase the cognitive load on developers during interpretation, integration, and secure deployment tasks. As a consequence, the probability of implementation errors, misconfiguration, insecure usage patterns, and unsafe integration practices increases, particularly in production environments where development constraints are more pronounced \cite{toruan2026security}. As a result, implementation quality becomes strongly dependent on individual developer expertise rather than being supported by robust system design. These limitations act as amplification mechanisms within the PQC implementation environment, where design shortcomings in tooling shape developer behaviour and increase the probability of security-relevant errors. This demonstrates that implementation risk is not solely derived from algorithmic complexity but also from the interaction between tool design and developer experience.

\begin{itemize}
    \item \textbf{Emerging and Latent Vulnerabilities:} \label{F_Cr_02}
\end{itemize}
    
Another critical cross-cutting challenge arises from the presence of emerging and latent vulnerabilities that may remain undetected during initial design and deployment phases \cite{bagirovs_applications_2024}. Although PQC algorithms are currently subject to extensive evaluation, their long-term security properties remain dependent on evolving cryptanalytic capabilities, implementation correctness, and parameter stability \cite{bagirovs_applications_2024}. As a result, previously unknown weaknesses may surface through advances in analysis techniques or shifts in adversarial capability. This uncertainty introduces temporal instability into PQC deployment environments, where security assumptions may degrade over time \cite{bagirovs_applications_2024}. Such degradation extends beyond cryptographic primitives and affects interconnected system components, including integration toolchains, deployment pipelines, and maintenance processes. As a result, vulnerabilities may emerge indirectly through system evolution rather than from immediately observable design defects. These delayed effects generate cascading impacts across implementation environments, where early design assumptions propagate into operational risks that become visible only under changing threat conditions. In practice, such latent weaknesses manifest across multiple dimensions of the implementation ecosystem. For example, at the organisational level (Section \ref{F_L04}), they contribute to delayed mitigation responses, weak governance adaptation, and ineffective planning. At the human level (Section \ref{F_L05}), they contribute to reduced awareness of evolving threats, limited preparedness for post-deployment changes, and decreased confidence in implementation decisions. Overall, latent vulnerabilities function as amplifiers of systemic risk, reinforcing cross-cutting propagation effects across the entire HOT dimensions. 

\section{Discussion}
\label{Discussion}

The synthesis identifies a difference between the implementation support represented in RQ1 and the challenges described in RQ2. The mapped approaches primarily provide technological resources and education, whereas the reported challenges also involve organisational planning, workforce preparation, and maintenance responsibilities (Sections \ref{RQ1_HOT} and \ref{RQ2_challenges}). The software engineering question arising from this difference concerns how specialised resources support the people and processes involved in implementation. We examine this question across the software lifecycle and introduce PQC-HOT as an analytical framework connecting implementation tasks with technical resources, developer capability, and organisational arrangements.

\subsection{Connecting Implementation Support Across the Software Lifecycle}
\label{Dis_RQ1}

The approaches identified in RQ1 provide complementary forms of support. For example, selection frameworks inform choices among algorithms and libraries \cite{decisionAlgo}, evaluation frameworks structure implementation assessment \cite{kannwischer_improving_2022, PQCLeo25}, and protocol integrations demonstrate PQC functionality within particular systems \cite{huelsing2021wireguard, aguilera_integrating_2024}. Together, these contributions identify resources for different engineering activities. Their value depends partly on the activity they address: a library supplies executable functionality, while a testing framework supplies mechanisms for examining selected implementation properties.

The relationship between these resources becomes important when developers move from one activity to another. The tooling and integration findings describe difficulties involving interoperability, documentation, and existing software dependencies \cite{hekkala_implementing_2023, aydeger_towards_2024}. Together with RQ1, these findings motivate examining whether the requirements guiding algorithm selection remain clear during integration, and whether later validation checks that those requirements are met. We interpret this as a coordination question concerning the continuity of decisions and assurance evidence across implementation activities.

This interpretation directs attention to the connections between specialised approaches. A useful investigation would trace which assumptions a resource establishes, which decisions it leaves to practitioners, and what evidence the next activity requires. For example, developers who select a library for its performance and compatibility should check these criteria again during integration. This analysis could identify where support is sufficient and where developers must address gaps themselves. The current mapping offers a a basis for this investigation without assuming that all specialised approaches are disconnected or incomplete.  

The organisational distribution adds a further consideration. Table \ref{sok_synthesis_hot} contains no approach classified primarily as organisational, although some provide organisational support. For example, PQCIP combines professional education, practical activities, and strategic planning \cite{PQCIP}. Organisational concerns also appear in the identified challenges. This raises a question for software engineering research: how do development teams connect implementation resources with assigned responsibilities, review processes, and ongoing support?

\subsection{Implementation Assurance and Developer Support}
\label{Dis_RQ2}

The implementation-security findings highlight the need to check whether software meets the security requirements of the cryptographic design. The reviewed studies report risks arising from cryptographic transformations, coding choices, and runtime behaviour \cite{howe_sok_2021}. Other evidence also shows that implementation errors and incorrect use can undermine cryptographic software \cite{crptoImmerr}. Teams therefore need evidence that their chosen software configuration works securely under its intended operating conditions.

PQClean illustrates this emphasis through how systematic software quality checks can support cryptographic standardisation \cite{kannwischer_improving_2022}. Its contribution supports integrating implementation checks into development processes. However, historical findings apply to the implementations and period studied. The enduring engineering concern is the distinction between selecting a standardised mechanism and validating its implementation within their application.

Integration also requires teams to assess compatibility, architectural dependencies, and operational constraints \cite{aydeger_towards_2024, giron_migrating_2023, tsili_scalable_2025}. They motivate evaluating the selected configuration within its intended workload and environment. Correctness, performance, and interoperability each answer a different engineering question. Therefore, teams should state what each assessment demonstrates and where evidence remains missing. Changes to components or security assumptions subsequently create reasons to revisit that assessment \cite{bagirovs_applications_2024,giron_migrating_2023}.

\label{Dis_HumanSupport}
Developer support helps practitioners produce and interpret evidence about their implementations. Toruan et al. \cite{toruan2026security} examine how developers use PQC APIs and documentation and identify opportunities to improve terminology, guidance, and workflow examples. These findings suggest that interfaces should provide guidance suited to users’ expertise. Providing functionality alone may leave developers unsure how to use it correctly, assessing usability therefore contributes to understanding the practical conditions of secure integration.

Education also supports implementation. Courses identifies challenges related to prior knowledge and technical depth  \cite{borrelli_designing_2024}, while professional training and laboratory activities offer practical learning opportunities  \cite{PQCIP,PQCLabEdu}. These findings motivate matching learning objectives to practitioner responsibilities. Understanding concepts, completing guided exercises, and independently checking an unfamiliar integration require different assessments. Transfer between them remains an empirical question that implementation-oriented education should address.

Organisational arrangements connect these capabilities to development practice. The reported workforce and resource challenges  \cite{aydeger_towards_2024,PQCIP} motivate examining whether practitioners have the support required for their assigned work. For a particular implementation, this includes questions about access to specialist review, authority to resolve uncertainties, and ownership of subsequent updates. Considering these arrangements alongside tools and education provides the basis for the PQC-HOT framework.

\subsection{PQC-HOT: An Analytical Framework for Software Implementation}
\label{PQC_HOT}

PQC-HOT contextualises the Human-Technology-Organisation interaction perspective \cite{Berglund2020HTOA} for the approaches and challenges synthesised in this review. Its analytical contribution is to connect the support functions identified in RQ1 with the requirements identified in RQ2 at the level of an implementation task. This task-based application asks what technical resources are available, what practitioners must understand and perform, and what organisational arrangements support completion and review. The framework provides a structure for examining correspondence between these elements.

\subsubsection{Dimensions and Evidence-Informed Relationships}

\begin{figure*}
  \includegraphics[width=\textwidth, height=0.4 \textheight ]{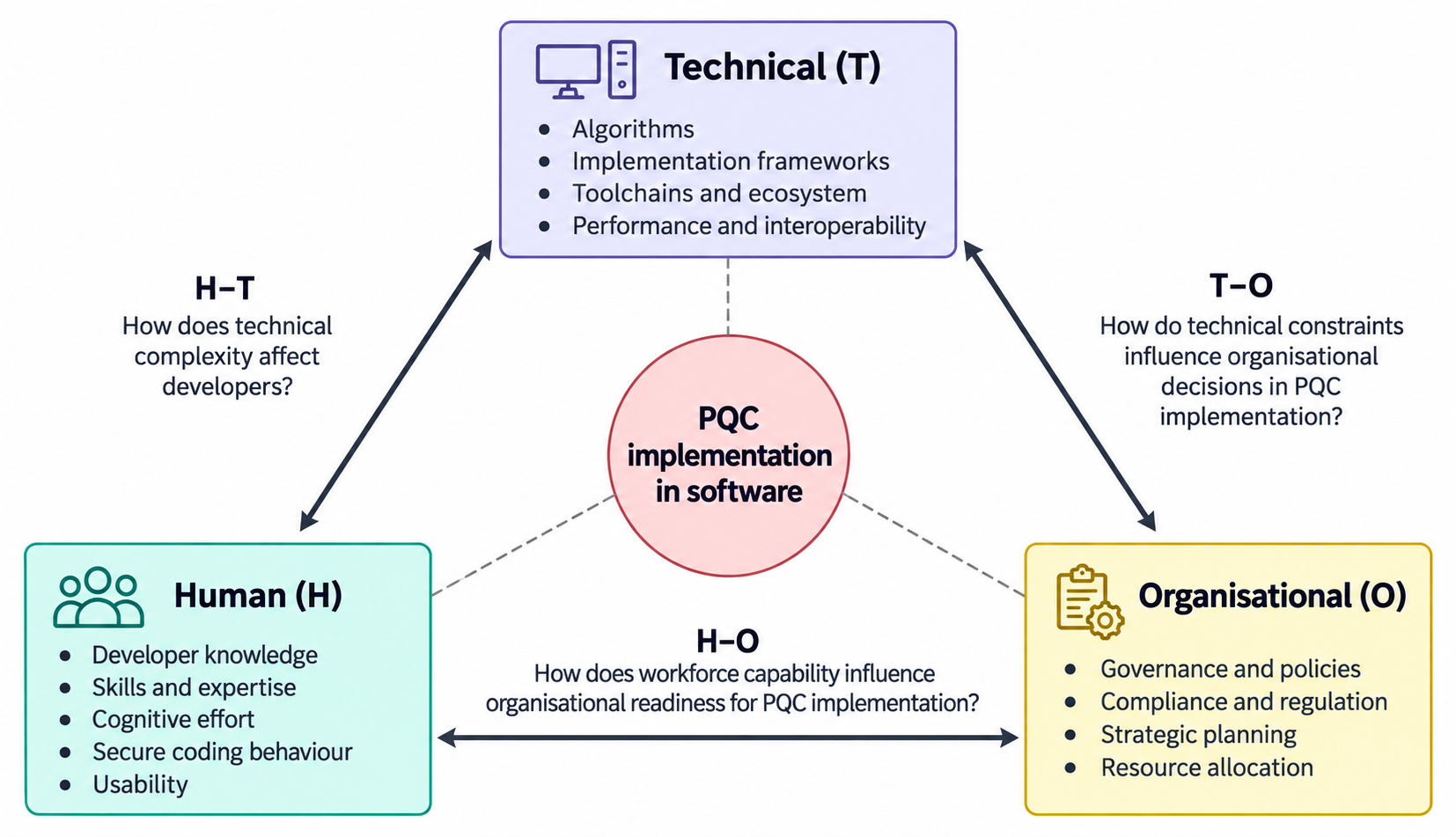}
  \caption{PQC-HOT: an analytical framework connecting human, organisational, and technological considerations in PQC implementation. The connections indicate relationships proposed for investigation.}
  \label{PQC_HOT_TriangleImage}
\end{figure*}

Figure \ref{PQC_HOT_TriangleImage} presents three dimensions. The \textit{Technological} dimension includes cryptographic mechanisms, software implementations, APIs, architectures, and validation resources. The \textit{Human} dimension includes role-specific knowledge, implementation practices, interpretation of guidance, and verification skills. The \textit{Organisational} dimension includes responsibilities, resources, review processes, operational requirements, and maintenance arrangements.

Three relationships follow from the synthesis. The \textit{Technological - Human} relationship connects API and documentation concerns to the decisions expected of developers \cite{toruan2026security, hekkala_implementing_2023}. It directs analysis towards whether interfaces and guidance communicate the requirements needed for correct use. The \textit{Technological - Organisational} relationship connects integration and resource constraints to planning and operational processes \cite{aydeger_towards_2024, giron_migrating_2023}. It raises questions about how organisations allocate responsibility for compatibility assessment, validation, and updates. The \textit{Human - Organisational} relationship connects workforce preparation to assigned responsibilities and available support \cite{aydeger_towards_2024,PQCIP}. It prompts examination of whether relevant expertise, training, and review opportunities are available when required.

Within PQC-HOT, \textit{alignment} means that the resources, capabilities, and organisational arrangements available for a task correspond to its implementation requirements. This definition permits examination of a concrete gap, such as a validation requirement for which neither sufficient expertise nor review ownership has been identified. The framework is an evidence-informed conceptual contribution: the cited findings motivate its relationships, while their causal significance and the framework's practical effectiveness require empirical evaluation.

\subsubsection{Illustrative Application to Library Integration}
\label{HOT_Example}

Consider a hypothetical development team integrating a PQC library into an existing application. The team must select an appropriate implementation, use its interface correctly, evaluate the integration, and manage subsequent changes. Table \ref{tabHOTApplication} illustrates how PQC-HOT can organise questions across these activities. The example demonstrates a proposed application of the framework; it does not report an observed project or additional review findings.

\begin{table*}[t]
\centering
\caption{Illustrative PQC-HOT analysis of a team integrating a PQC library into an existing application. Entries are proposed analytical questions.}
\label{tabHOTApplication}
\small
\setlength{\tabcolsep}{4pt}
\renewcommand{\arraystretch}{1.15}
\begin{tabularx}{\textwidth}{@{}p{0.14\textwidth} >{\raggedright\arraybackslash}X >{\raggedright\arraybackslash}X >{\raggedright\arraybackslash}X@{}}
\toprule
\textbf{Activity} & \textbf{Technological} & \textbf{Human} & \textbf{Organisational} \\
\midrule
Selection & Which functionality, compatibility, and resource requirements apply? & Can the team interpret and compare the selection criteria? & Who reviews and approves the selection? \\
Integration & What interface requirements and failure conditions must be handled? & Can developers apply the guidance and recognise incorrect use? & Is specialist review available for unresolved questions? \\
Validation & Which checks address the required implementation and system properties? & Can practitioners interpret results and investigate failures? & Who defines acceptance criteria and reviews the evidence? \\
Maintenance & Which changes require reassessment or regression checks? & Can the team evaluate the implications of an update? & Who monitors dependencies and owns implementation changes? \\
\bottomrule
\end{tabularx}
\end{table*}

Suppose the team has access to a testing tool but has not assigned responsibility for interpreting its output. Recording the tool's availability captures the technological resource. Examining the same validation task across the other dimensions reveals questions about reviewer competence and decision ownership. The proposed response would address the identified requirement by assigning a suitably prepared reviewer and defining how unresolved findings are handled. The example demonstrates how the framework can organise an examination of support for a task, while leaving the actual adequacy of that support to evidence from the project.

The analysis can also follow decisions between activities. Requirements recorded during selection can inform integration review and validation criteria; maintenance ownership can then be linked to the configuration and evidence produced. This application makes the relationships between dimensions concrete by associating them with work products, decisions, and responsibilities.

\subsubsection{Intended Applications}
\label{propositions}

PQC-HOT has four intended uses. For \textit{analysis}, it structures examination of the support required for a defined implementation task. For \textit{comparison}, it characterises the primary and secondary contributions of approaches to that task. For \textit{gap identification}, it helps analysts look for requirements without corresponding resources, capabilities, or responsibilities. For \textit{method development}, it informs the design of processes connecting implementation activities with evidence and ownership. These uses require explicit task requirements and coverage evidence; primary HOT classification alone cannot establish a support gap.

\subsection{Implications for Software Engineering Research and Practice}
\label{Res_Im}

\subsubsection{Research Priorities}

\textit{Study continuity across implementation activities:} The complementary resources identified in RQ1 and the integration difficulties reported in RQ2 motivate studies that follow development work across selection, integration, validation, and updating \cite{decisionAlgo, PQCLeo25, aydeger_towards_2024}. Researchers can trace how requirements are recorded and reused, where assumptions become unclear, and what additional support practitioners seek. Such studies would provide direct evidence about coordination between implementation resources.

\textit{Evaluate developer support through task performance:} API usability findings motivate assessing guidance and interface design against observable outcomes \cite{toruan2026security}. Relevant measures include correct usage, recognition of errors, and verification behaviour. Educational evaluations can similarly distinguish knowledge acquisition from independent integration and debugging performance \cite{borrelli_designing_2024, PQCIP, PQCLabEdu}. Unfamiliar tasks and follow-up assessments would help examine transfer and retention. These evaluation designs would test the capabilities needed for implementation rather than infer them from participation alone.

\textit{Investigate organisational ownership of implementation assurance:} The organisational challenges in RQ2 motivate examining who approves implementation choices, reviews validation evidence, and maintains dependencies \cite{aydeger_towards_2024, giron_migrating_2023}. Case studies can compare stated responsibilities with observed practice and identify requirements lacking resources or ownership. Readiness assessments could then be developed from these observations and evaluated against meaningful project outcomes.

\textit{Evaluate the analytical value of PQC-HOT:} Framework evaluation should examine whether analysts can apply its categories consistently and whether the task-based relationships reveal actionable issues beyond separate technical, human, and organisational checklists. Comparative and longitudinal studies can subsequently investigate whether changes addressing identified gaps improve implementation outcomes. Such evaluation should distinguish benefits attributable to individual changes from benefits associated with coordinated support.

\subsubsection{Engineering Implications}

The implementation security findings motivate making assurance requirements explicit when selecting and integrating PQC components \cite{howe_sok_2021, kannwischer_improving_2022}. For each configuration, teams can record the relevant requirements, identify the checks that address them, and assign responsibility for reviewing the resulting evidence. This produces a concrete connection between an implementation decision and the basis for accepting it.

Developer support should match the decisions exposed by the interface and the responsibilities assigned to its users. The usability evidence motivates clearer terminology and workflow examples \cite{toruan2026security}; the educational findings motivate practical assessment appropriate to the intended role \cite{PQCIP, PQCLabEdu}. An application developer integrating a library and a specialist modifying cryptographic internals may require different guidance and review expertise. Making these responsibilities explicit helps define the support each task requires.

Compatibility and update requirements should likewise be connected to testing and maintenance ownership \cite{aydeger_towards_2024, giron_migrating_2023}. The illustrative application shows how teams can use PQC-HOT to examine these connections before accepting an integration or subsequent change. Its purpose is to structure engineering decisions and expose questions requiring evidence, with effectiveness to be assessed in the intended setting.

\subsubsection{Scope of the Interpretation}

The discussion characterises the reviewed evidence and derives analytical and design implications from it. Primary classifications do not capture every contribution of an approach, and publication counts do not measure effectiveness or industrial adoption. The challenge synthesis identifies concerns and motivates possible relationships without ranking their causal importance. These boundaries support the use of PQC-HOT as a structured basis for investigation while preserving the distinction between reported findings, proposed applications, and demonstrated outcomes.

\section{Conclusion}
\label{conclu}

This study systematised the literature on PQC implementation in software systems through the Human, Organisational, and Technological (HOT) perspective. It identified implementation approaches, synthesised reported challenges, and examined the relationship between available support and implementation requirements.

RQ1 (Section \ref{RQ1_Intervention}) identified guidelines, frameworks, tools and libraries, and educational interventions. Technological support received greater representation in the extracted mapping, while no approach was classified primarily as organisational. Nevertheless, some approaches contributed to organisational preparation through secondary functions. RQ2 (Section \ref{RQ2_challenges}) identified challenges spanning implementation security, system integration, tooling, organisational governance, and developer capability. Together, these findings reveal a contrast between the primary support functions of the mapped approaches and the breadth of the reported implementation challenges. 

Building on this synthesis, we proposed PQC-HOT as an evidence-informed analytical framework connecting implementation tasks with technical resources, developer capabilities, and organisational arrangements. The framework provides a structure for analysing approaches, comparing their contributions, examining support gaps, and informing method development. Its illustrative application demonstrates how these considerations can be examined across library selection, integration, validation, and maintenance. The framework's practical effectiveness and the relationships it proposes require empirical evaluation.

The study highlights the importance of connecting cryptographic mechanisms with the engineering activities, practitioner support, and organisational responsibilities involved in their use. Future research should investigate these connections through developer studies, industrial case studies, and longitudinal evaluations. Such work can assess how coordinated support contributes to secure and maintainable PQC implementation in software systems.

\printcredits

\section*{Funding Sources}
This research did not receive any specific grant from funding agencies in the public, commercial, or not-for-profit sectors.

\bibliographystyle{model1-num-names}

\bibliography{reference_CnS_Sok}

\end{document}